\documentclass[aps,prl,twocolumn,superscriptaddress]{revtex4-1}
\usepackage{amsmath,amssymb,amsthm,graphics,graphicx,bm,subfigure}
\usepackage{graphicx}
\usepackage{xcolor}
\usepackage{hyperref}
\usepackage{listings}
\hypersetup{
    unicode=false,          % non-Latin characters in Acrobat‚Äôs bookmarks
    pdftoolbar=true,        % show Acrobat‚Äôs toolbar?
    pdfmenubar=true,        % show Acrobat‚Äôs menu?
    pdffitwindow=false,     % window fit to page when opened
    pdfstartview={FitH},    % fits the width of the page to the window
    pdfauthor={Mario Arnolfo Ciampini},     % author
    colorlinks=true,       % false: boxed links; true: colored links
    linkcolor=blue,          % color of internal links
    citecolor=red,        % color of links to bibliography
}

\graphicspath{{Figures/}}

\begin{document}

\definecolor{dkgreen}{rgb}{0,0.6,0}
\definecolor{gray}{rgb}{0.5,0.5,0.5}
\definecolor{mauve}{rgb}{0.58,0,0.82}

\lstset{frame=tb,
  	language=Matlab,
  	aboveskip=3mm,
  	belowskip=3mm,
  	showstringspaces=false,
  	columns=flexible,
  	basicstyle={\small\ttfamily},
  	numbers=none,
  	numberstyle=\tiny\color{gray},
 	keywordstyle=\color{blue},
	commentstyle=\color{dkgreen},
  	stringstyle=\color{mauve},
  	breaklines=true,
  	breakatwhitespace=true
  	tabsize=3
}

\newcommand{\tzz}{TEM$_{00}\,$}
\newcommand{\tzo}{TEM$_{10}\,$}
\newcommand{\er}[1]{\textcolor{purple}{#1}}

\title{Experimental nonequilibrium memory erasure beyond Landauer's bound}
\author{Mario A. Ciampini}
\affiliation{Vienna Center for Quantum Science and Technology (VCQ), Faculty of Physics, University of Vienna, A-1090 Vienna, Austria}
\author{Tobias Wenzl}
\affiliation{Vienna Center for Quantum Science and Technology (VCQ), Faculty of Physics, University of Vienna, A-1090 Vienna, Austria}
\author{Michael Konopik}
\affiliation{Institute for Theoretical Physics I, University of Stuttgart, D-70550 Stuttgart, Germany}
\author{Gregor Thalhammer}
\affiliation{Division for Biomedical Physics, Medical University of Innsbruck, 6020 Innsbruck, Austria}
\author{Markus Aspelmeyer}
\affiliation{Vienna Center for Quantum Science and Technology (VCQ), Faculty of Physics, University of Vienna, A-1090 Vienna, Austria}
\affiliation{Institute for Quantum Optics and Quantum Information (IQOQI), Austrian Academy of Sciences, 1090 Vienna, Austria}
\author{Eric Lutz}
\affiliation{Institute for Theoretical Physics I, University of Stuttgart, D-70550 Stuttgart, Germany}
\author{Nikolai Kiesel}
\affiliation{Vienna Center for Quantum Science and Technology (VCQ), Faculty of Physics, University of Vienna, A-1090 Vienna, Austria}\date{\today}

\begin{abstract}
The clean world of digital information is based on noisy physical devices. Landauer's principle provides a deep connection between information processing and the underlying thermodynamics by setting a lower limit on the energy consumption and heat production of logically irreversible transformations. While Landauer's original formulation assumes equilibrium, real devices often do operate far from equilibrium.
We show experimentally that the nonequilibrium character of a memory state enables full erasure with reduced power consumption as well as negative heat production. 
We implement the optimized erasure protocols in an optomechanical two-state memory. To this end, we introduce dynamical shaping of nonlinear potential landscapes as a powerful tool for levitodynamics as well as the investigation of far-from-equilibrium processes.
\end{abstract}

\maketitle

\begin{figure}[t]
\centering
\includegraphics[width=\columnwidth]{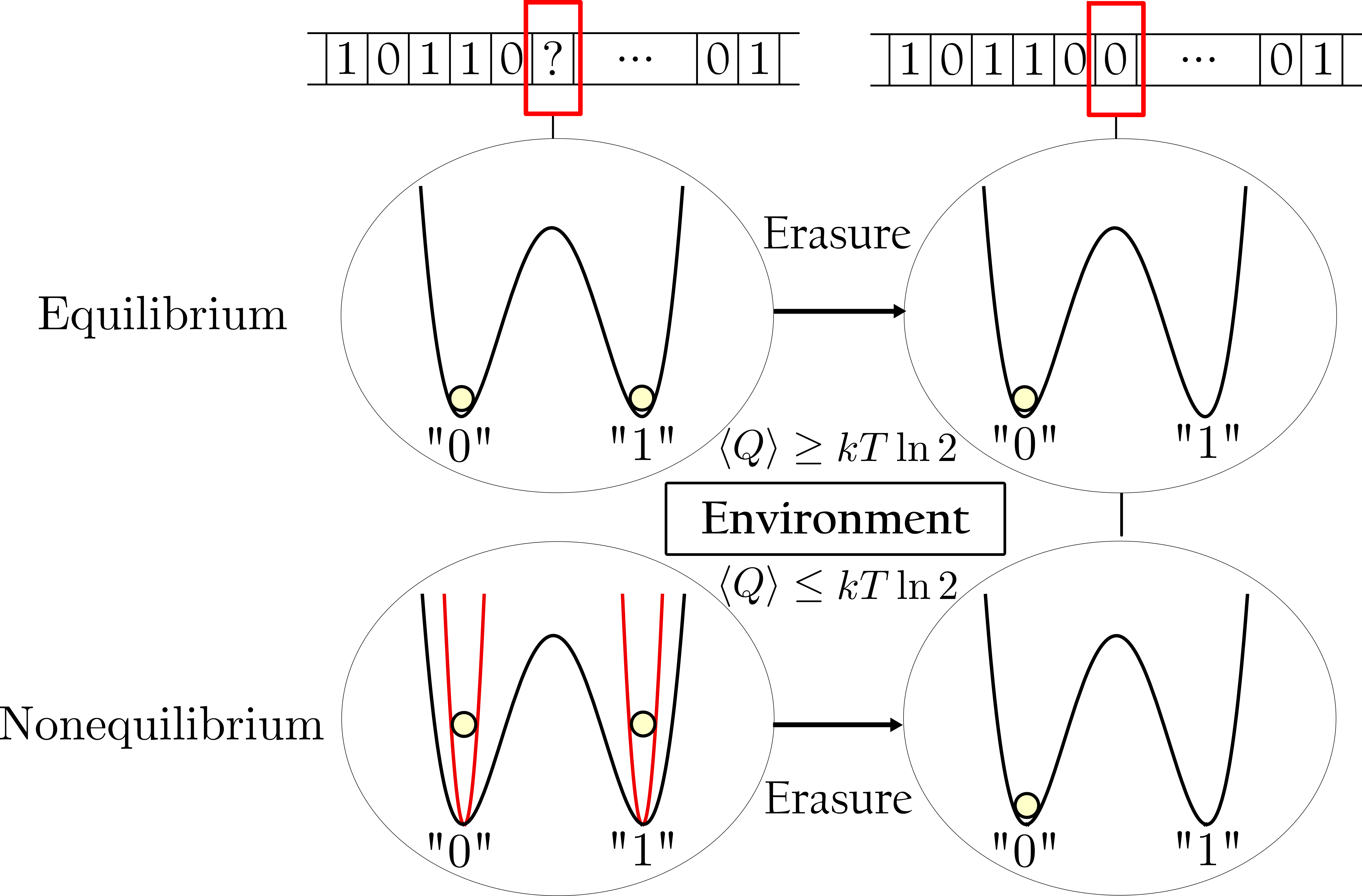}
\caption{\textbf{Nonequilibrium information erasure}. One bit of information stored in a generic double-well memory may be erased by resetting the system to state "0", irrespective of the initial state "0" or "1". While dissipated heat is always larger than $kT\ln2$ when information is stored in an equilibrium  state, it might be smaller when energy and entropy of a nonequilibrium memory state are properly harnessed.}
\label{fig:ConceptualScheme}
\end{figure}

Heat production poses fundamental challenges in computer hardware development. From a fundamental perspective, dissipation in logically irreversible operations is tightly connected to Landauer's bound \cite{Landauer1961}, a central result of information thermodynamics \cite{ben82,dissipation2,thermoinfo1,thermoinfo2}.
Specifically, Landauer's principle, in its original equilibrium formulation, states that the erasure of a single bit of information  in an environment at temperature $T$ (Fig.~\ref{fig:ConceptualScheme}) consumes at least $k T \ln2$ of work and entails the same amount of dissipated heat ($k$ denotes the Boltzmann constant) \cite{Landauer1961,ben82,dissipation2,thermoinfo1,thermoinfo2}. Today, real computational devices operate several orders of magnitude above this value \cite{Pop2010}, which they may reach, however, already in a few decades  \cite{the17}. 

%Real devices, however, usually operate far-from-equilibrium as emphasis is given on fast switching as much as energy consumption \cite{Ercan2013}. Today, practical devices operate at the very best a factor of 1000 above Landauer's bound \cite{Pop2010}. Furthermore, it has been argued that a system subject to thermal stochastic noise is incapable of reaching Landauer’s bound \cite{Norton2012}.

Meanwhile, proof-of-principle experiments with colloidal particles and nanomagnetic systems have already achieved irreversible information erasure with an energy dissipation close to the Landauer limit \cite{Berut2012,Orlov2012,Jun2014,Hong2016}. Notably, Landauer's bound has been reached for near-equilibrium erasure times on the order of minutes for overdamped colloidal particles trapped in a double-well potential \cite{Berut2012, Jun2014} and recently in only 100ms with an underdamped cantilever \cite{Ciliberto21}. Beyond such confirmations, the necessity to extend Landauer's principle has also been discussed for finite success probability \cite{gam13,Talukdar2018}, finite time \cite{Proesmans2020, Konopik2021} as well as for asymmetric initial conditions in near-equilibrium protocols, both theoretically \cite{sag09,dil10} and experimentally \cite{Gavrilov2016}. 
%, Lambson2011=theory; The nanomagnet experiment by Martini is experimentally a factor >10 above Landauer, Orlov show quite more than just erasure, but also are quite above Landauer (> x10)
 Real devices, however, usually do not operate near equilibrium. Information in modern (volatile) memories is, for example, commonly stored in nonequilibrium states,  whose preparation requires a given amount of energy and entropy \cite{sta15}. Their information is lost when power is switched off. Recent theoretical results suggest that far-from-equilibrium physics may have a significant impact on the thermodynamic bounds of the erasure process \cite{esp11,Klaers2019,Konopik2018}. In particular, work consumption and heat dissipation may be controlled and reduced below the $k T \ln2$ limit when information is stored in a nonequilibrium state \cite{Konopik2018}.
%as emphasis is given on fast switching as much as energy consumption
%Recently, a generalized Landauer's bound for memory erasure with far-from-equilibrium initial conditions has been introduced theoretically . It predicts that a system that is sufficiently far from equilibrium can use this property  to avoid dissipation of heat and to reduce the work required during an erasure process. 
%Given the need for speed in actual processing devices, the question is whether additional bounds for memories that operate far-from-equilibrium exist, as well as potential trade-offs in the quality of memory operations, speed, heat dissipation, and character of the storage device in question

%We introduce dynamically programmed optical levitation of anharmonic potential as a powerful tool to assess such questions, extending the toolbox of levitated optomechanics \cite{Askin1971, Gieseler2018, Rondin2017}.

\begin{figure*}[h!tb]
\centering
\includegraphics[width=0.9\textwidth]{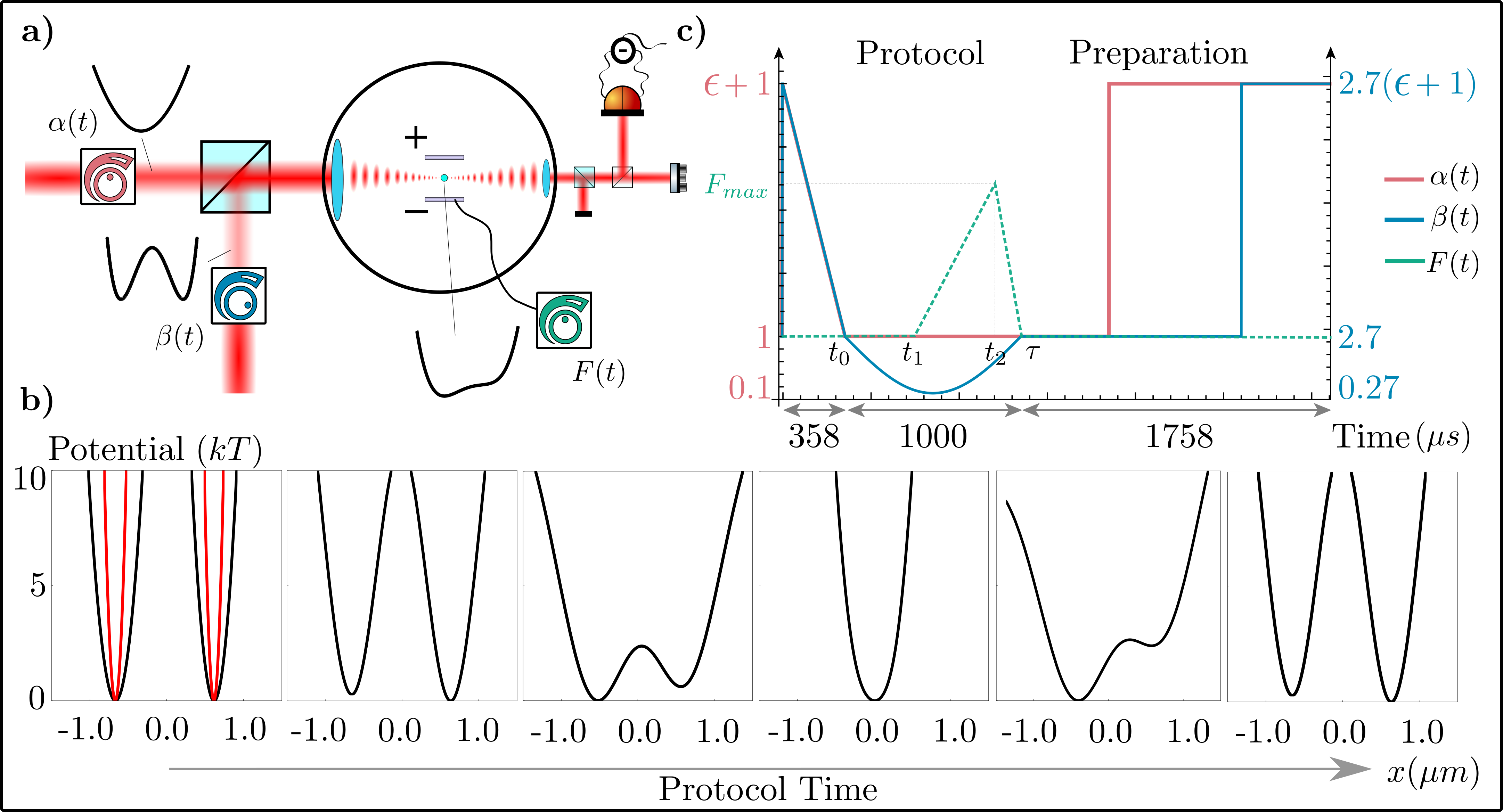}
\caption{\textbf{a) Schematic experimental setup:} Two laser beams, one in a horizontally polarized \tzz mode and the other in a vertically polarized \tzo mode, are combined at a polarizing beam-splitter (PBS) and tightly focused inside a vacuum chamber to create an optical trap. The power of the two laser beams, $\alpha(t)$ and $\beta(t)$, can be varied independently by acousto-optical modulators (AOMs) to modulate the shape of the double-well potential. Electrodes are placed in the vicinity of the optical trap (distance $100 \mu$m) to control the tilt $F(t)$ of the potential. After the vacuum chamber, the vertically polarized \tzo mode is separated with a PBS. The major part of the \tzz mode is reflected back into the tweezer, creating a standing wave trap. A small fraction of the light from the optical trap is fed into a standard split mirror detection to measure the  position of the charged nanoparticle in the radial direction. \textbf{b) Potential snapshots  during reset.} Experimentally reconstructed trapping potentials for the nonequilibrium parameter $\epsilon=4$ at different times during information reset. The red curve shows the potential for an initial equilibrium state ($\epsilon = 0$). \textbf{c) Nonequilibrium reset protocol:} Time-dependence  of the three control parameters $\alpha(t)$, $\beta(t)$ and $F(t)$ for an optimal nonequilibrium reset protocol. The $y$-axes are not to scale. }
\label{fig:ExpSetup}
\end{figure*}

We here experimentally verify that the nonequilibrium  character of an initial memory state is a useful resource for information reset in a general double-well potential. We concretely analyze the erasure of a bit stored in a nonequilibrium state of an optomechanical two-state memory. We consider erasure cycles  consisting of a  preparation and a reset phase. We demonstrate a reduction of both  consumed work and dissipated heat during reset far below the equilibrium value of $k T \ln2$, even reaching negative values, in agreement with a generalized nonequilibrium Landauer bound \cite{Konopik2018}. To this end, we introduce dynamically programmed control of anharmonic potentials for optical levitation as a powerful and versatile tool to modulate nonlinear potentials on fast timescales in the underdamped regime. With this, we further extend the toolbox of levitated optomechanics \cite{Askin1971,Tongcang2013,Gieseler2013,Rondin2017,Gieseler2018, Mil21},  where the dynamic control was limited to amplitude modulation of Gaussian potentials and forces.\\

We first prepare a nonequilibrium initial state to store one bit of information in a double-well potential using a levitated silica nanosphere  confined in an optical trap created by two laser beams (Fig.~\ref{fig:ExpSetup}a).  This generic model for a one-bit memory encodes state "0" ("1") when the particle is in the left (right) potential well. Information is erased by resetting the memory to state "0" irrespective of the initial state \cite{Landauer1961,ben82,dissipation2,thermoinfo1,thermoinfo2}. We  implement this reset-to-zero by modulating the shape of the potential by varying the laser power, decreasing the barrier height and applying a tilt to the left (Fig.~\ref{fig:ExpSetup}bc). We design and optimize the reset  protocol to minimize applied work and dissipated heat, using temporal and spatial control over the system parameters. We specifically exploit optical levitation to
control the strength of the underdamped coupling of the nanoparticle to its environment, achieving a reduction of the reset time by 5 orders of magnitude compared to overdamped experiments \cite{Berut2012, Jun2014, Gavrilov2016}. 
%We find agreement with a generalized Landauer's bound for non-equilibrium initial states that we have derived recently \cite{Konopik2018}. We also show that the dissipated heat can even become negative for initial states sufficiently far from equilibrium. In other words, the erasure is not heating but effectively cooling the environment in our experiment.

%Besides, optical levitation enables tuning of the dissipation time scale in the system. We exploit this property to operate in a critically damped regime where the speed of the protocol on the order of milliseconds, improving by 5 orders of magnitude over previous experiments in colloidal systems.

%\section{Results}

%We consider a particle in a double-well potential as a generic model for a one-bit memory (Fig. \ref{fig:ConceptualScheme}). Initially, an arbitrary state of the memory is prepared, with the particle in the right (left) well encoding a bit “0” (“1”). The goal of the erasure protocol is to reset the memory state to “0” in a way that is independent of the initial state. Qualitatively, this is achieved by a reduction of the barrier height followed by a tilt of the potential,  followed by the reintroduction of the barrier (Fig. \ref{fig:ExpSetup}c).  To minimize dissipation, optimized temporal and spatial control over the system parameters is required (Fig. \ref{fig:ExpSetup}b). Recently, such optimal protocols for initial symmetric non-equilibrium states have been introduced \cite{Konopik2018}, which we implement here experimentally.

We consider an erasure protocol made  of a preparation stage, during which the nonequilibrium memory  is initialized, and a reset stage, during which the memory is erased. The preparation step is absent for equilibrium memories as originally considered by Landauer \cite{Landauer1961,ben82,dissipation2,thermoinfo1,thermoinfo2}. Because of the initial out-of-equilibrium state of the double-well system, Landauer's principle needs to be extended \cite{esp11}. Applying the laws of thermodynamics to this nonequilibrium configuration, a generalized Landauer bound holds for the mean dissipated heat $\langle Q \rangle$ and the mean consumed work $\langle W \rangle$ in  the reset phase \cite{Konopik2018},
\begin{align}
\langle Q \rangle & \geq Q_L = T \Delta I -\Delta U_\text{res},\\
\langle W \rangle &\geq W_L = T\Delta I, 
\label{eq:workheat}
\end{align}
where $\Delta U_\text{res}$ is the variation of internal energy of the system during reset of duration $\tau$ and $\Delta I$  is the change of relative entropy, $I(t)=k \int \rho(t)\ln\left[\rho(t)/(\rho_\text{eq}(t))\right] dx dv$, an entropic distance between a nonequilibrium memory state $\rho(x,v,t)$ and the corresponding equilibrium state $\rho_\text{eq}(x,v,t)$ ($x$ is the position and $v$ the velocity) \cite{sch80}. The brackets $\langle \cdot\rangle$ denote an ensemble average over many repetition of the  process. Equations (1) and (2)  reduce to the standard Landauer limits for initial and final equilibrium states that correspond to $\Delta I = k\ln2$ and $\Delta U_\text{res} = 0$. We note that  the nonequilibrium Landauer bounds for heat and work, $Q_L$ and $W_L$, may be controlled through the initial entropic distance to equilibrium $I(0)$ and the nonequilibrium preparation energy, $\Delta U_\text{pre} = -\Delta U_\text{res}$. 
%
%It has also been demonstrated that this allows to use non-equilibrium properties of initial states to control the dissipated heat and applied work, giving rise to a generalized Landauer's bound \cite{Konopik2018}. This bound shows that average dissipated heat and average work applied can be reduced. By how much depends on the relative entropy, 
%$I(t)=k_B \int \rho(t)ln\left[\rho(t)/(\rho_{eq}(t))\right] dx dp$,  between the initial and final states $\rho(t)$ and the  equilibrium state $\rho_{eq}(t)$, respectively:

%Here Q is the heat dissipated during the protocol, W is the work performed to erase the memory, $T=300K$ is the temperature of the environment. %and $H(t_0)=-\int dx \rho_{eq} (x,t_0) log_2 [\rho_{eq} (x,t_0)]$ is the Shannon entropy of the initial and final distribution. 
%The energy required to prepare the initial out-of-equilibrium state is given by $-dU_{res}$ and $\langle\rangle$ denotes an ensemble average over many repetition of the protocol. The above relations show that by controlling the preparation of the state it is possible to control $I(0)$ and $\Delta U_{res}$ to reduce both the heat and work required to reset the memory below Landauer’s bound, which takes the well-known value of $k_B T ln2$ for a symmetric memory \cite{Konopik2018}. Initial potential asymmetry and non-complete reset reduce the variation in entropy during the protocol, further reducing the expected energy consumption during the reset protocol \cite{Gavrilov2016} (see Supplementary Information).

\begin{figure}[t!]
\centering
\includegraphics[width=.89\columnwidth]{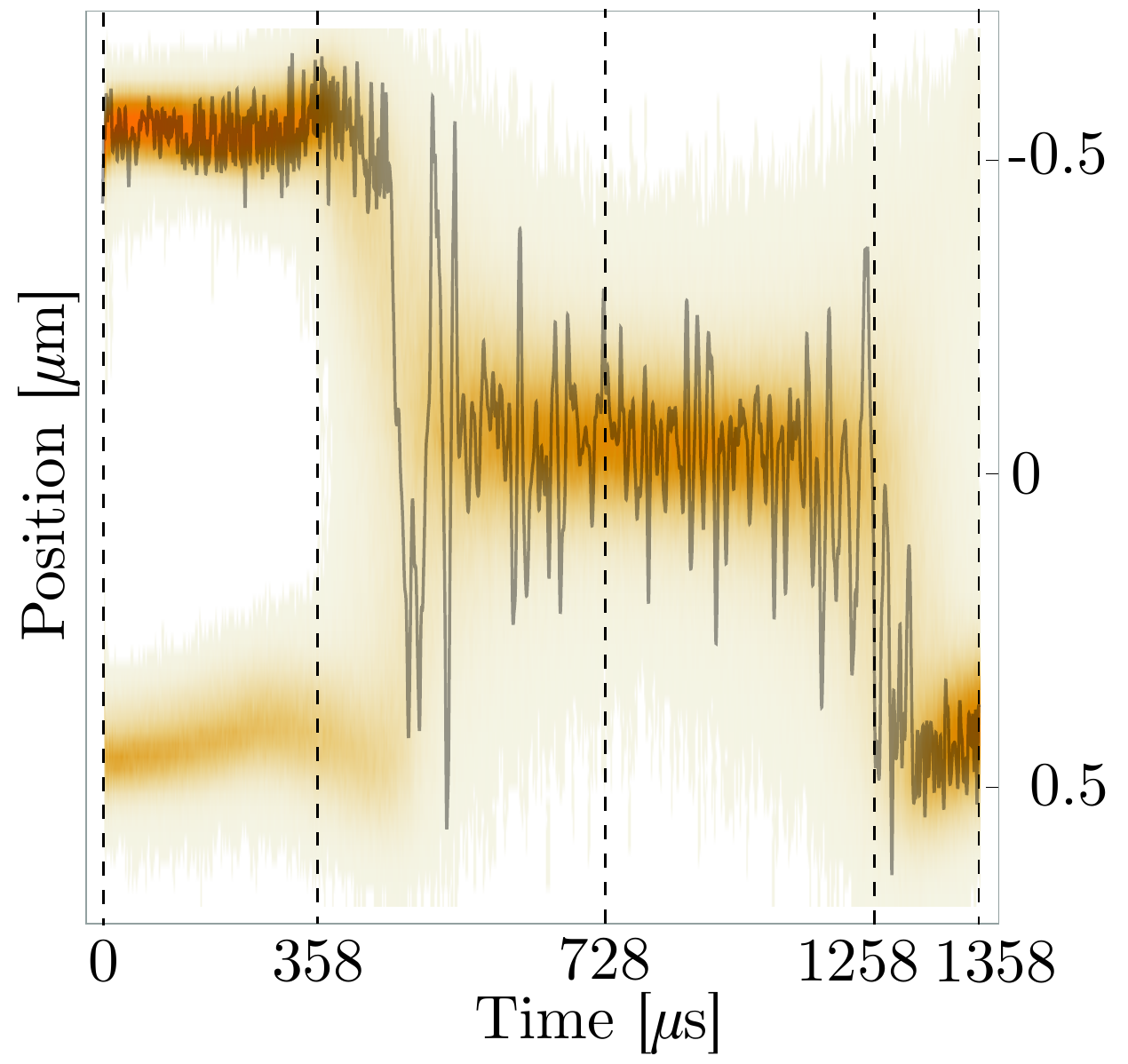}
\caption{\textbf{Position distribution during  reset.} Exemplary single stochastic trajectory of a nanoparticle (black line) from state "1" to state "0" measured during a nonequilibrium reset protocol  for $\epsilon=4$.  The corresponding position density for about 20.000 repetitions of the protocol is represented by the orange cloud. The reset protocol lasts $\tau=1358\mu$s. $t_0$=358$\mu s$ represents the start of the equilibrium erasure, $t_1$=728$\mu s$ identifies the introduction of the external force, which reaches the maximum magnitude $F_{max}$ at $t_2$=1258$\mu s$.} \label{fig:Timetrace}
\end{figure}

In our experiment, we realize a dynamical double-well optical trap with the configuration shown in Fig. \ref{fig:ExpSetup}a. A charged silica particle (radius = 74nm) is confined in an optical tweezer inside a vacuum chamber. The pressure $P=(30.0\pm0.3)$mbar is set to create  $\Gamma=(35.0\pm0.2)$kHz of viscous damping. The optical trap consists of two tightly focused (waist $W\approx0.6\mu$m and wavelength $\lambda =1064$nm) and orthogonally polarized lasers with spatial modes $\text{TEM}_{00}$ and $\text{TEM}_{10}$ (Methods). We control their respective  powers, $\alpha(t)$ and $\beta(t)$,  independently  using acousto-optic modulators (AOMs). Depending on their power ratio, a harmonic, quartic or double-well trap with tunable barrier height is created transversely to the optical path along the $x$-axis (Fig.~\ref{fig:ExpSetup}b). Our setup allows to treat the motion along this axis independently of the other directions (Methods).  We further tilt the potential for the charged nanoparticle with a well-controlled force $F(t)$ generated by two electrode blades separated by 100 micrometers along the $x$-axis, mounted far outside the optical trap (Supplementary Information). The resulting potential is of the form $V(x,t)= -[\alpha(t)a+\beta(t)b/2 x^2] \exp[-(cx^2)/2]+F(t)x$, where $a$, $b$ and $c$ are geometric parameters that depend on the  setup (Supplementary Information).

We resolve the dynamics $x(t)$ of the nanoparticle in the double-well potential using a split detection of the TEM$_{00}$ mode in transmission \cite{gie12}. Due to the time-variation in the potential and its anharmonic shape, the data covers the frequency range from DC to 1 MHz. We perform the thermodynamic analysis of the data by evaluating the work, $W=\int_0^\tau\, dt \partial V(x,t)/\partial t$, along individual trajectories $x(t)$ \cite{sek10}, after filtering the resulting spectrum with a low-pass filter at 0.28MHz, which also removes residuals of the harmonic $y$- and $z$-motion (Methods). We additionally determine the dissipated heat $Q$ via the first law of thermodynamics (Methods).  We calculate their respective average values by repeating each measurements about 20.000 times.

%
%This setting allows to experimentally investigate the produced heat and applied work during erasure protocols applied to memory states with a controlled non-equilibrium character. 

In order to put our  results into  perspective, we first 
 implement the reset protocol for a memory initialized in an equilibrium state \cite{Berut2012}, $\rho(0) = \exp(-\beta V)/Z$, where $Z$ is the partition function and $\beta=1/(kT)$ the inverse temperature. In contrast to  overdamped experiments \cite{Berut2012, Jun2014, Gavrilov2016}, we operate at significantly reduced damping rates and thus much shorter reset times ($\tau = 1000\mu$s). 
% For sufficient statistics, we perform N=21082 runs of this protocol, including a re-initialization step. Here, $\alpha_{eq}$, $\beta_{eq}$ determine the initial potential.
 We determine the symmetry of the initial equilibrium position distribution (labeled by the parameter $\epsilon=0$) by reconstructing the potential from the measured trajectories (Supplementary Information). We find the probability  $P_{\epsilon=0}^{0}(0)=36.6\%$ ($P_{\epsilon=0}^{1}(0)=63.4\%$) for the nanoparticle to be in state "0" ("1"). We reset  the memory to state "0"  with a success probability $P_{\epsilon=0}^{0}(\tau)=99.6\%$. The particle hence ends up in the unwanted state "1" in only $0.4\%$ of all    runs. We obtain the  mean applied work $\langle W \rangle_\text{eq}=(0.583\pm 0.046) k T$ and the mean dissipated heat $\langle Q \rangle_\text{eq}=(0.620\pm0.064) k T$. These values are consistent with the theoretical expectations for our asymmetric memory: given the measured values of  the probabilities $P_{\epsilon=0}^{0}(0)$  and $P_{\epsilon=0}^{0}(\tau)$, the equilibrium Landauer bounds for work and heat are  indeed  $\langle W \rangle_{L}^\text{eq}=\langle Q \rangle_{L}^\text{eq}=(0.57\pm 0.10) k T$ (Supplementary Information). 

We next consider the reset of a nonequilibrium memory. We create the out-of-equilibrium initial state of the double-well potential by letting the particle equilibrate in a steep preparation potential given by $\alpha_\text{pre}=(\epsilon+1)\alpha$ and $\beta_\text{pre}=(\epsilon+1) \beta$ (Fig. \ref{fig:ExpSetup}b, red line). These parameters are chosen in order to decrease energy and entropy of the initial nonequilibrium state as compared to the equilibrium state given by $\epsilon =0$. In particular, the nonequilibrium state is narrower than the corresponding equilibrium state.  The value of $\epsilon$ thus controls the departure from equilibrium  of the  initial memory  state.  We design and optimize the reset protocol in order to properly harness  the preparation energy and entropy, and therefore  minimize consumed  work and dissipated heat \cite{Konopik2018}. Figure \ref{fig:ExpSetup}c shows the  protocol used  for the fast dynamical control of the three parameters $\alpha(t)$, $\beta(t)$ and $F(t)$ for an optimal information reset followed by a reinitialization of the nonequilibrium state.
Figure \ref{fig:Timetrace} displays the measured position distribution of the particle (red)  over time  for $\epsilon=4$, along with one exemplary stochastic trajectory (black) showing a reset from bit "1" $(x<0)$ to bit "0" $(x>0)$.  The initial asymmetry in this case is  $P_{\epsilon=4}^{0}(0)=56.7\%$ $(P_{\epsilon=4}^{1}(0)=43.3\%)$ for the potential well corresponding to state "0" ("1"). We reset the memory to state "0" in time $\tau = 1358\mu$s with a success probability of $P_{\epsilon=4}^1(\tau)=97.9\%$. 

%We determine the heat by the First Law $Q_i=\Delta E_i-W_i$, where we evaluated the change in total internal energy $\Delta E_i$ between the beginning and the end of protocol i. For non-equilibrium states up to $\epsilon=4$, we show in Fig. \ref{fig:exp_result}a the applied average work $\langle W \rangle$ and dissipated heat $\langle Q \rangle$ as a function of the non-equilibrium parameter $\epsilon$. Explicit values for reset probabilities and work/heat calculations are provided in the Supplementary Materials. 
%Our results are consistent with the theoretical expectations given by the generalized Landauer's bound.

Figure 4 represents the averaged consumed work $\langle W \rangle$ (blue bars) and dissipated heat $\langle Q \rangle$ (red bars) for various nonequilibrium parameters $\epsilon$. We first observe that these nonequilibrium quantities  are smaller than the corresponding equilibrium values (dashed bars), taking the asymmetry of each initial state into account. We further clearly see that  both $\langle W \rangle$ and $\langle Q \rangle$ decrease with increasing departure from equilibrium, in  full agreement with Eqs.~(1) and (2) (white bars): the consumed work is significantly reduced to   $\langle W \rangle_{\epsilon=4}= 0.053\pm0.021 kT$, while the dissipated heat even becomes negative  for $\epsilon \gtrsim 2.5$, reaching $\langle Q \rangle_{\epsilon=4}= -0.393\pm0.032 kT$, indicating that heat is actually absorbed during information reset. Explicit values for reset probabilities and work/heat evaluations are provided in the Supplementary Information.

\begin{figure}[!t]
\centering
\includegraphics[width=\columnwidth]{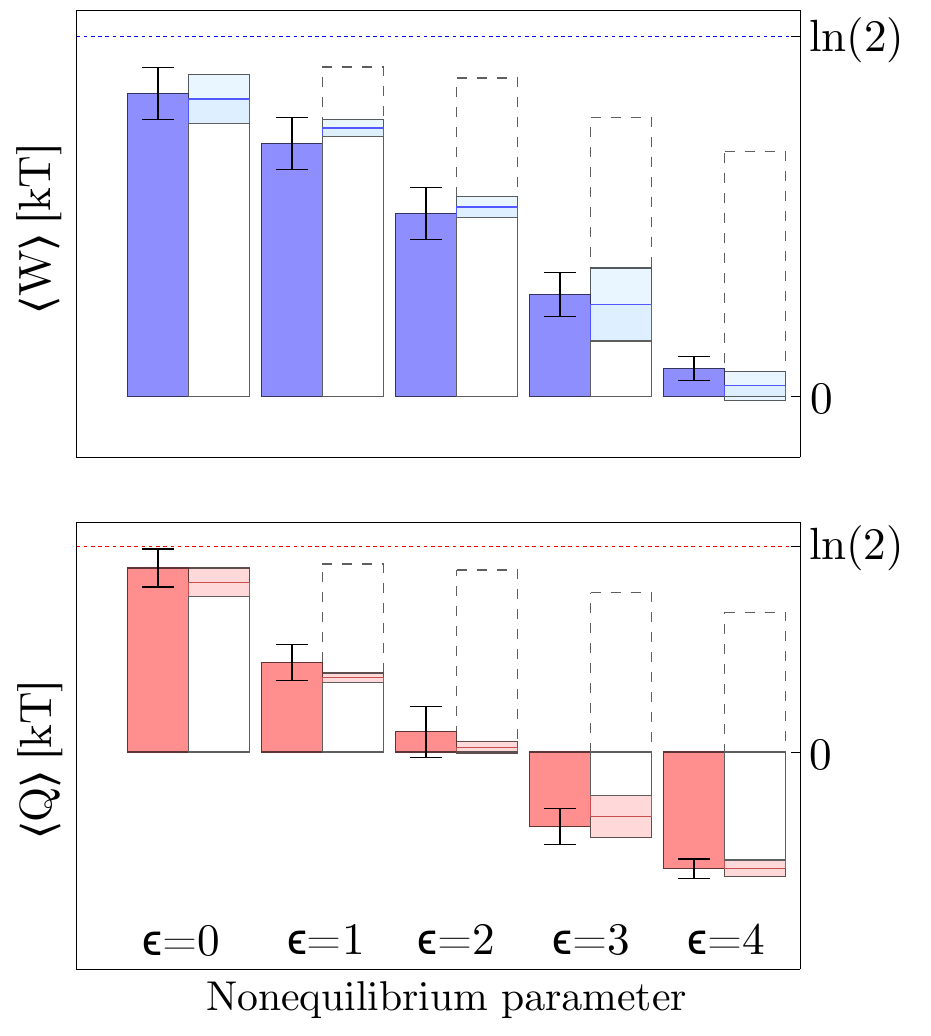}
\caption{\textbf{Energetics of nonequilibrium  reset.} Average consumed work (blue bars) and dissipated heat (red bars) for various nonequilibrium  parameters $\epsilon$. Error bars are 1-sigma confidence intervals obtained for approximately 20.000 repetitions. We observe a significant suppression of both quantities as $\epsilon$ is increased, in agreement with the generalized Landauer bounds (1) and (2) (solid white bars). 
The shaded areas represent the uncertainty on this value introduced by the experimentally determined asymmetry of the corresponding double-well potentials. The dashed white bars show Landauer's bound for the same configuration but an equilibrium initial state.
The dashed horizontal lines indicate the $kT \ln2$ limit valid for  equilibrium symmetric memories. 
}
\label{fig:exp_result}
\end{figure}

%We observe a reduction in heat and work with increasing non-equilibrium parameter $\epsilon$. These values are reduced further by a small correction (dashed bars in Fig. \ref{fig:exp_result}a) accounting for the asymmetry of the potential and imperfect success rate, consistently with Eq. \ref{eq:workheat} (see Supplementary Information for a full analysis). %Note that the main contribution in the reduction compared to the standard Landauer bound (kTln2) is due to the non-equilibrium initial state.
%For $\epsilon>\approx 2.5$, the average dissipated heat switches sign, in other words the erasure protocol effectively results in cooling the environment. %Note that this cooling is possible a consequence of the work that is needed to prepare the initial non-equilibrium memory state. 

To conclude, we have experimentally demonstrated  the usefulness of nonequilibrium memory states to achieve information reset below the equilibrium limit of $kT\ln2$ joule, both for work consumption and heat dissipation, in agreement with a generalized nonequilibrium Landauer principle. Such reduction is physically possible when entropy and energy of the nonequilibrium memory state preparation is properly harnessed by the reset protocol. Clearly, there is no way to avoid the overall thermodynamic cost during memory erasure as enforced by the second law for a full cycle. Yet, the existence of distinct preparation and reset stages for nonequilibrium memories offers remarkable thermodynamic flexibility. The cost of erasure can thus be shifted from the usual reset phase to the nonequilibrium preparation phase, as we have shown here. This property offers new means to control the thermodynamics of logically irreversible operations \cite{Wolpert2019}, for example in proposed computer architectures where preparation and processing zones are spatially separated \cite{Kish2016}. They may hence be implemented locally at no energetic cost and without heat dissipation or even entail cooling of the local environment.
Our  results were enabled by novel dynamically programmed optical micromanipulation of a nanosphere trapped in a nonlinear potential and levitated in mild vacuum. This is an essential step towards bringing the same level of virtuous control known from optical micromanipulation of overdamped colloids in liquid (e.g. \cite{Grier2002}) to levitated systems in vacuum. Dynamically programmable optical levitation is not only a versatile platform to investigate information thermodynamics  but, together with the preparation of pure quantum states of nanoparticle motion \cite{Tebbenjohanns2020, Delic2020, Magrini2021}, may also become a powerful tool for quantum state engineering beyond Gaussian states and thus a novel approach to matter-wave interferometry in macroscopic quantum physics.
\\

\noindent \textbf{Methods}\\
We experimentally create the double-well potential by the combination of orthogonally polarized TEM$_{00}$(H) and TEM$_{10}$(V) modes. The latter is obtained by using a Spatial Light Modulator (SLM).  At equilibrium, we have a power of 50mW in the vacuum chamber in the harmonic mode, at the start of the protocol, and approximately 135mW in the TEM$_{10}$ mode. This is the lowest possible configuration in power where we get a stable particle and a barrier high enough to prevent spontaneous jumping between the two wells. The nonequilibrium initial state is created by maintaining the same power proportion between the two beams and increasing the power according to $\epsilon$. For example, the value $\epsilon=4$ requires having power 250mW in H and  675mW in V.  
We resolve the fast dynamics along the $x$ direction (up to a particle frequency of  $2\pi \cdot 250$ kHz) in the potential using a split detection of the TEM$_{00}$ mode in transmission \cite{gie12}. In order to allow for a 1D treatment of the experiment, the harmonic motion along the $y$- and $z$-axes need to significantly exceed the motional frequencies along the $x$-axis. This allows for spectral filtering of the motion along these directions. While this is intrinsically reached along the $y$-axis, we increase the frequency along the $z$-axis with a standing wave configuration, where a part of the TEM$_{00}$ mode is back-reflected into the focus, essentially creating discs of light. Full reconstruction of the trapping potential $V(x,t)$ is given in the Supplementary Material.

We evaluate the work along single trajectories $x(t)$ by summing over the discrete changes in the potential energy for incremental time steps of $dt=0.16\mu$s. We determine the heat via the first law  as $\Delta E = E(\tau)-E(0) = W+Q = m[v(\tau)^2-v(0)^2)/2+[V(x(\tau),\tau)-V(x(0),0)]$.  Detector and force calibration account for position-dependent sensitivity as discussed in the Supplementary Information. 
\\

\noindent \textbf{Acknowledgements}\\
We thank Lorenzo Magrini for useful discussions and Monika Ritsch-Marte for her support.  This work was supported by the Austrian Science Fund (FWF): Y 952-N36, START. We further acknowledge financial support from the German Science Foundation (DFG) under project FOR 2724 and the QuantERA project TheBlinQC (Project No. 864032; via the EC, the Austrian ministries BMDW and BMBWF and research promotion agency FFG). M.A.C. acknowledges support from the FWF Lise Meitner Fellowhip (M2915, "Entropy generation in nonlinear levitated optomechanics").

\bibliographystyle{naturemag}
\bibliography{references}

\end{document}

\newcommand{\tzz}{$\text{TEM}_{00} $}
\newcommand{\tzo}{$\text{TEM}_{10} $}

\lstset{frame=tb,
  	language=Matlab,
  	aboveskip=3mm,
  	belowskip=3mm,
  	showstringspaces=false,
  	columns=flexible,
  	basicstyle={\small\ttfamily},
  	numbers=none,
  	numberstyle=\tiny\color{gray},
 	keywordstyle=\color{blue},
	commentstyle=\color{dkgreen},
  	stringstyle=\color{mauve},
  	breaklines=true,
  	breakatwhitespace=true
  	tabsize=3
}

\title{Supplementary Information: Experimental nonequilibrium memory erasure beyond Landauer's bound}
\author{Mario A. Ciampini}
\affiliation{Vienna Center for Quantum Science and Technology (VCQ), Faculty of Physics, University of Vienna, A-1090 Vienna, Austria}
\author{Tobias Wenzl}
\affiliation{Vienna Center for Quantum Science and Technology (VCQ), Faculty of Physics, University of Vienna, A-1090 Vienna, Austria}
\author{Michael Konopik}
\affiliation{Institute for Theoretical Physics I, University of Stuttgart, D-70550 Stuttgart, Germany}
\author{Gregor Thalhammer}
\affiliation{Division for Biomedical Physics, Medical University of Innsbruck, 6020 Innsbruck, Austria}
\author{Markus Aspelmeyer}
\affiliation{Vienna Center for Quantum Science and Technology (VCQ), Faculty of Physics, University of Vienna, A-1090 Vienna, Austria}
\affiliation{Institute for Quantum Optics and Quantum Information (IQOQI), Austrian Academy of Sciences, 1090 Vienna, Austria}
\author{Eric Lutz}
\affiliation{Institute for Theoretical Physics I, University of Stuttgart, D-70550 Stuttgart, Germany}
\author{Nikolai Kiesel}
\affiliation{Vienna Center for Quantum Science and Technology (VCQ), Faculty of Physics, University of Vienna, A-1090 Vienna, Austria}\date{\today}
\maketitle

\section{Generalised Landauer's Bound}
We reformulate Landauer's principle in terms of relative entropies between distributions, to take into account the non-equilibrium characteristic of the initial state. Starting from the First and Second law of thermodynamics, we can rewrite work as:
\begin{equation}
\langle W\rangle \geq \Delta U - T\Delta S \equiv W_L.
\end{equation}
Where $\Delta U$ is the change in internal energy and $\Delta S$ is the change in entropy. The time-dependent equilibrium Hamiltonian can be rewritten using the equilibrium distribution $H_{eq} = -kT \ln Z_{eq} -kT \ln \rho_{eq}.$ Therefore:
\begin{eqnarray}
W_L &=& \langle H_{eq}\rangle_i - \langle H_{eq}\rangle_f +T(S_f -S_i)\\ \nonumber
&=& kT (\langle \ln \rho_{eq}\rangle_i - \langle \ln \rho_{eq}\rangle_f +\langle \ln \rho_f\rangle_f -\langle \ln \rho_i\rangle_i )\\ \nonumber
&=& T ( I(\rho_f,\rho_{eq}) -  I(\rho_i,\rho_{eq}))=T \Delta I,
\label{eq:genbound}
\end{eqnarray}
with the relative entropy being defined as $I(\rho_{f/i},\rho_\text{eq})=k \langle \ln\left[\rho_{f/i}/\rho_{eq}\right]\rangle=k \int \rho_{f/i}\ln\left[\rho_{f/i}/\rho_\text{eq}\right] dx dv$. Heat is then given by 
\begin{equation}
\langle Q \rangle \geq  W_L - \Delta U \equiv Q_L,
\label{eq:3}
\end{equation}
using the sign convention $\Delta U = W-Q$.

\subsection{Asymmetry corrections}
Here we derive Landauer's bound by taking into account asymmetry in the potential and imperfect erasure.
We approximate the double-well potential with high-barrier by fitting both wells with harmonic oscillator potentials, for the theoretical bound we also ignore any potential momentum distribution differences, i.e. we assume that the momentum distribution stays approximately thermal. 
There are three relevant distributions, the equilibrium distribution $\rho_\text{eq}$ which is used to determine the equilibrium potential, the initial distribution $\rho_\text{i}$ and the final one $\rho_\text{f}$. 

The position distribution for the equilibrium distribution is given by:
\begin{equation}
\rho_{eq}= 
\begin{cases}
p^0_{eq}  e^{-\beta \sigma^0_{eq} (x-x_0)^2}/Z^0_{eq}  & \text{for } x>0\\
p^1_{eq}   e^{-\beta \sigma^1_{eq} (x+x_1)^2}/Z^1_{eq} & \text{for } x<0.
\end{cases}
\end{equation}
The distribution parameters $p^{0/1}_{eq},\sigma^{0/1}_{eq}$ are replaced by $p^{0/1}_{i},\sigma^{0/1}_{i}$ when describing the initial out-of-equilibrium distribution,  and by $p^{0/1}_{f},\sigma^{0/1}_{f}$ for the final one. Here, we assume for simplicity that the position of the minimum does not change before/after reset. $p^{0/1}$ represents the probability of the particle being on either side of the wells, thus $p^{0}=1-p^{1}$ holds. In order to recover the well-known Landauer's bound of $kT \ln2$, $p^{0}_{eq}=p^{0}_{i}=1/2$ and $p^{0}_{f}=1$. This treatment allows to recover bounds accounting for asymmetry in the potential and a non-perfect reset. We do assume, however, that $p^{0/1}_{eq}=p^{0/1}_{i}$: we improve the reset by solely altering the entropy within the wells.

In thermal equilibrium, we can reconstruct the potential by using $V(x)=-\text{kT} \ln(p(x))$. 
%We can back-trace the potential offset given by the parameters $p^{0/1}_{eq}$ by considering $V^{0/1}_{eq}=-\text{kT} \ln p^{0/1}_{eq}$. Expanding up to the second order around $x_1$ and $x_2$ the position of the two wells, this results in the potential:
%\begin{equation}
%V_{eq}\approx \begin{cases}
%\sigma^0_{eq}(x-x_1)^2 + \alpha_0  & \text{for } x>0\\
%\sigma^1_{eq}(x+x_2)^2 + \alpha_1  & \text{for } x<0.
%\end{cases}
%\end{equation}
%Landauer's bound follows from the First and Second law of thermodynamics:
%\begin{align}
%W_{L}&=\Delta U  - T \Delta S = \nonumber \\ 
%&=\langle H^{eq}\rangle_f - \langle H^{eq}\rangle_i + k_B T \langle \ln \rho_f \rangle_f - k_B T \langle \ln \rho_i \rangle_i. %\label{eq:landauergen}
%\end{align}
Calculating explicitly the quantities in Eq.\ref{eq:genbound}, the energies are given by 
\begin{equation}
\langle H_{eq}\rangle_{i}=\frac{kT}{2} \left(p^0_{i} \frac{\sigma^0_{eq}}{\sigma^0_{i}} + p^1_{i} \frac{\sigma^1_{eq}}{\sigma^1_{i}}\right)-kT(p^0_{i}\ln p^{0}_{eq} + p^1_{i}\ln p^{1}_{eq}),
\end{equation}
for the initial case and replacing $p^{0/1}_{f}$ for $p^{0/1}_{i}$ as well as $\sigma^{0/1}_{f}$ for $\sigma^{0/1}_{i}$ for the final case.
The entropy difference is given by
\begin{eqnarray*}
\Delta S &=& -k\left(\int \rho_f \ln \rho_f dx - \int \rho_i \ln \rho_i dx \right)\\
&=& k(p^0_{i} \ln p^0_{i} + p^1_{i}\ln p^1_{i} - p^0_{f} \ln p^0_{f} - p^1_{f} \ln p^1_{f}  \\
&&+p^0_{f}\ln Z^0_f + p^1_{f}\ln Z^1_f -p^0_{i}\ln Z^0_{i} - p^1_{i}\ln Z^1_{i})\\
&=&k \ln \frac{{(p^0_{i})}^{p^0_{i}} {(p^1_{i})}^{p^1_{i}}}{{(p^0_{f})}^{p^0_{f}} {(p^1_{f})}^{p^1_{f}}} + k\ln\frac{{(Z^0_f)}^{p^0_{f}} {(Z^1_f)}^{p^1_{f}}}{{(Z^0_{i})}^{p^0_{i}} {(Z^1_{i})}^{p^1_{i}}}.
\end{eqnarray*}
Using Eq. \ref{eq:genbound}, the expression for the generalized Landauer's bound for our potential is: 

\begin{widetext}
\begin{eqnarray}
\frac{W_L}{kT}&=& \frac{1}{2} \left( \frac{\sigma^0_{eq}(p^0_{i}\sigma^0_{i}-p^0_{i}\sigma^0_f)}{\sigma^0_{i} \sigma^0_f}+ \frac{\sigma^1_{eq}(p^1_{i}\sigma^1_{i}-p^1_{i}\sigma^1_f)}{\sigma^1_{i} \sigma^1_f}\right)\\  \nonumber
&&+\ln\frac{(p^0_{eq})^{p^0_{i}} (p^1_{eq})^{p^1_{i}}}{(p^0_{eq})^{p^0_f} (p^1_{eq})^{p^1_f}} - \ln \frac{(p^0_{i})^{p^0_{i}} (p^1_{i})^{p^1_{i}}}{(p^0_{f})^{p^0_{f}} (p^1_{f})^{p^1_{f}}} - \ln\frac{(Z^0_f)^{p^0_f} (Z^1_f)^{p^1_f}}{(Z^0_{i})^{p^0_{i}} (Z^1_{i})^{p^1_{i}}}\\ \nonumber
&=& \frac{1}{2} \left( \frac{\sigma^0_{eq}(p^0_{f}\sigma^0_{i}-p^0_{i}\sigma^0_{f})}{\sigma^0_{i}\sigma^0_{f}}+  \frac{\sigma^1_{eq}(p^1_{f}\sigma^1_{i}-p^1_{i}\sigma^1_{f})}{\sigma^1_{i}\sigma^1_{f}}\right)\\ \nonumber
&&+\ln\frac{ (p^0_{f})^{p^0_{f}} (p^1_{f})^{p^1_{f}}}{(p^0_{eq})^{p^0_{i}} (p^1_{eq})^{p^1_{i}}}  + \frac{1}{2}\ln\frac{(\sigma^1_f)^{p^1_{f}} (\sigma^1_{f})^{p^1_{f}}}{(\sigma^0_{i})^{p^0_{i}} (\sigma^1_{i})^{p^1_{i}}},
\end{eqnarray}
\label{eq:landauergen2}
\end{widetext}
where in the last line we assumed $p^0_{eq}=p^0_{i}$. There are three contributions to the work cost. The latter two are comprised of the logical component given by the probabilities and the internal potential width components. %given by the partition sums of the respective wells. 
The first term is the energy difference between the initial locally squeezed distribution and the equilibrium one. The thermal energy $kT/2$ of the thermal distribution is reduced by the initial distribution in the equilibrium potential. 
The first fast step thus is accompanied with an energetic cost to work that makes the work larger than heat in the Landauer's bound (up to $k T /2$). If we ignore the fast step, then this term would vanish and work/heat would be equal in the symmetric case and only differ because of the potential offsets. As the first step is faster than the movement of the particle, this work offset is just added numerically when retrieving the bound.

We further  explicitly distinguish the contribution to the generalised Landauer's bound coming from the non-equilibrium character of the initial state ($\Delta W_{neq}$) from the one coming from the asymmetry of the potential ($\Delta W_{asymm}$). To this end, we rewrite Eq. \ref{eq:landauergen2} using $p^{1}=1-p^0$, $\sigma^0=\sigma^1$, $\sigma_{i}=\epsilon \sigma_{eq}$. We obtain:
\begin{widetext}
\begin{equation}
W_L/k_B T = \frac{1}{2} \left(2 \log \left((1-p^0_f)^{1-p^0_f} (p^0_f)^{p^0_f} (1-p^0_{eq})^{p^0_f-1} (p^0_{eq})^{p^0_f}\right)-\frac{1}{\epsilon }+\log \left(\frac{1}{\epsilon }\right)+1\right) = \ln 2 - (\Delta W_{asymm} + \Delta W_{neq})/k_B T.
\end{equation}
\label{eq:landauergen3}
\end{widetext}
In Table \ref{tab:data} we report the experimental probabilities, as well as the asymmetry correction to the bound given in Eq. \ref{eq:landauergen3}. The probability $p^0$ ($p^1$) is calculated as the frequency of the particle being in well "0" ("1") over the total number of repetitions. Errors on these parameters are evaluated by dividing the whole number of protocol repetition in M statistically independent subsets from which the standard deviation is calculated and propagated.

%In Tab. \ref{tab:data} we report $\Delta S_{asym}$, the correction to the symmetric Landauer's Bound due to the asymmetry of the original potential. Uncertainties are evaluated by dividing the dataset in 10 groups, and calculating the standard deviation over the resulting probabilities A, B, C, D, which are then propagated in the expression of the entropy.

%To summarize, the generalised Landauer's bound can be retrieved by merely fitting harmonic potentials to the experimental distributions, as long as the positions of the minima stay the same for all of the distributions and the equilibrium logical distribution of the particles in the double well is the same as for the initial non-equilibrium distribution ($A=C$).  Furthermore, the first fast process during the protocol has an 'artificial' work cost accompanied to it, that can just be added since the movement of the particles are not relevant in the time scale when the potential contraction happens. 

\section{Trapping scheme}
\begin{figure}[h!tb]
\centering
\includegraphics[width=0.90\columnwidth]{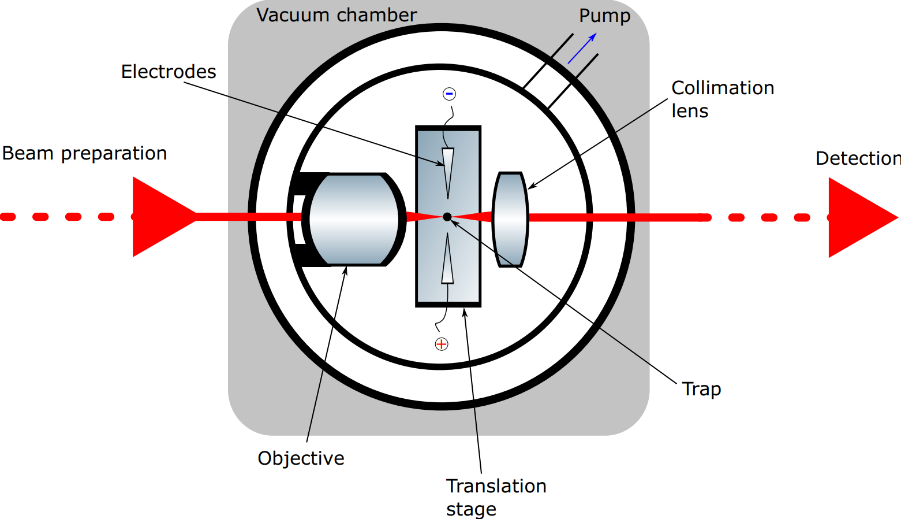}
\caption{\textbf{Schematic representation of the vacuum chamber}}
\label{fig:vacuumchamber}
\end{figure}

\begin{figure*}[h!tb]
\centering
\includegraphics[width=0.90\textwidth]{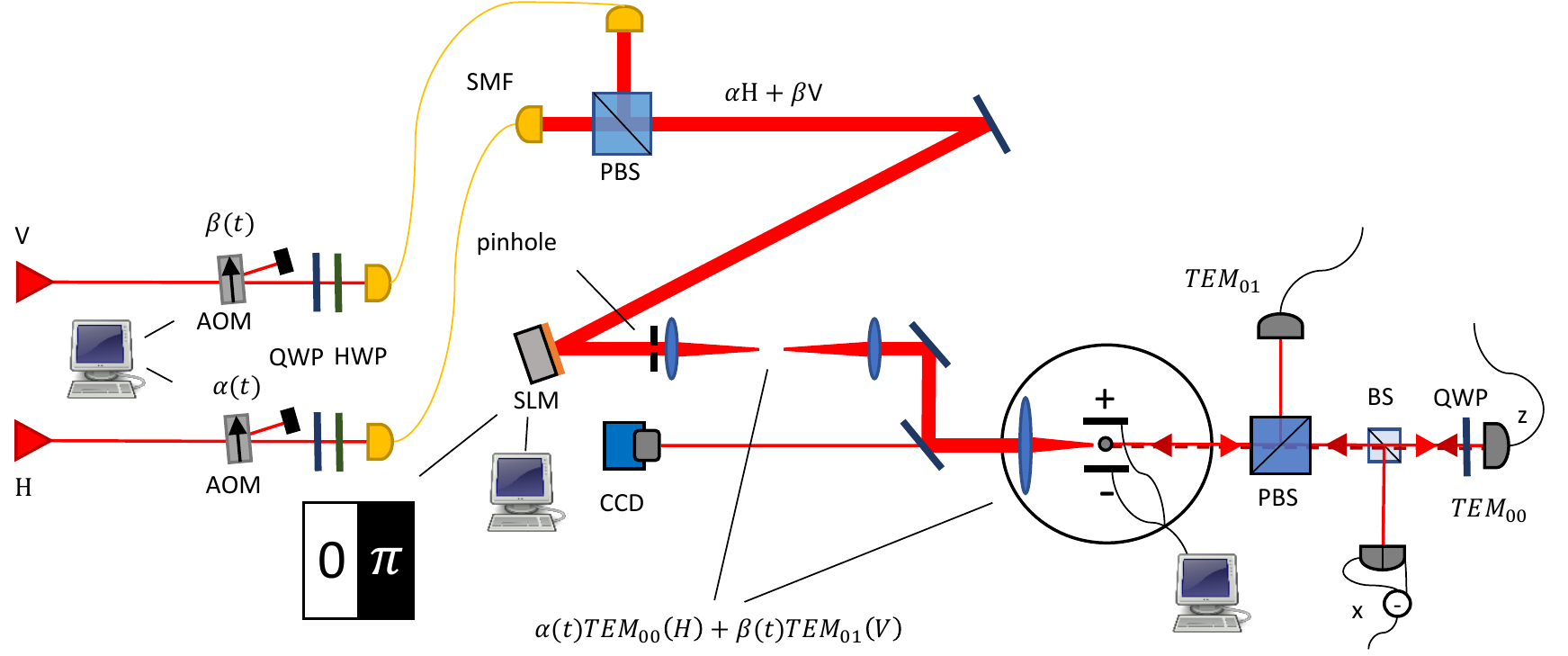}
\caption{\textbf{Experimental setup:} Two independent, orthogonal polarised laser beams (H: Mephisto (Coherent) + Keopsys Fiber Amplifier, V: Opus (Laser Quantum)) are modulated in intensity using independently controlled AOM's working in first-order diffraction and frequency shifted by 80MHz. Light is coupled into SMF and compensated in polarisation, then expanded to a beam diameter of approx 7mm and recombined into a PBS. An SLM is programmed to create a \tzo mode for the vertically polarised light. In the vacuum chamber a microscope objective (NA=0.8) creates the optical trap. Electrodes are placed around the focal point and can be positioned through a translation stage. The light from the trap is then collected and filtered in polarisation. Vertically polarised light is used to monitor the \tzo power, horizontally polarised light is split: 1\% of the light is used to measure the particle position using a split-mirror detection scheme, the remaining 99\% of the light is reflected into the vacuum chamber to create the standing wave. List of acronyms: PBS - polarizing beam splitter, AOM - acusto-optical modulator, SLM - spatial light modulator, SMF - single mode fiber, CCD - charged-coupled device fast camera, BS - beam splitter, HWP - half-wave plate, QWP - quarter-wave plate.}
\label{fig:expsetup}
\end{figure*}

We create the trap with a CFI TU Plan Fluor EPI 50x, NA=0.8, WD=1mm (Nikon Corp.) objective in the vacuum chamber is mounted. Between the objective and the collimation lens, two razor blades, spaced by $\approx 100 \mu m$, are mounted to create the necessary linear unbalancing force. The electrode holder is a 3D-printed custom designed element, made of Accura Xtreme White (Proto Labs GmbH). The spacing is small enough that the electrodes are fixed in their position. The holder is screwed onto a MX25 translation stage that is controlled by a CU30 controller (Mechonics AG.). This allows for a three-dimensional movement of the electrodes with 2 mm total travel distance in each direction (see Fig. \ref{fig:vacuumchamber}). Particles are trapped with the standard method using an ultrasonic nebulizer. We perform the experiment at a pressure in the vacuum chamber of $(30.0\pm0.3)$ mbar. The \tzo mode is created by means of a 512x512 spatial light modulator (SLM) (Meadowlark, Inc.), which has been previously calibrated to account for non-linear response of the pixels and non-flatness of the active surface. A simple splitting of the SLM active area to create a $\pi$ phase shift at the center of the beam (see Fig. \ref{fig:expsetup}) creates the wanted potential.    

\section{Calibration}
\subsection{Position Detection and Detector Calibration}

%\begin{figure}[h!tb]
%\centering
%\includegraphics[width=0.90\columnwidth]{tem10_scheme.pdf}
%\caption{\textbf{Representation of laser intensity.} \tzz (yellow) and \tzo (blue) laser intensity profile, with %same waist. When a particle is trapped in one of the wells, the \tzz light scattered by the particle and detected %is significantly reduced (red dashed line shows the intensity difference), leading to significantly lower %detection sensitivity.}
%\label{fig:tem01_scheme}
%\end{figure}

\begin{figure}[h!tb]
\centering
\includegraphics[width=0.93\columnwidth]{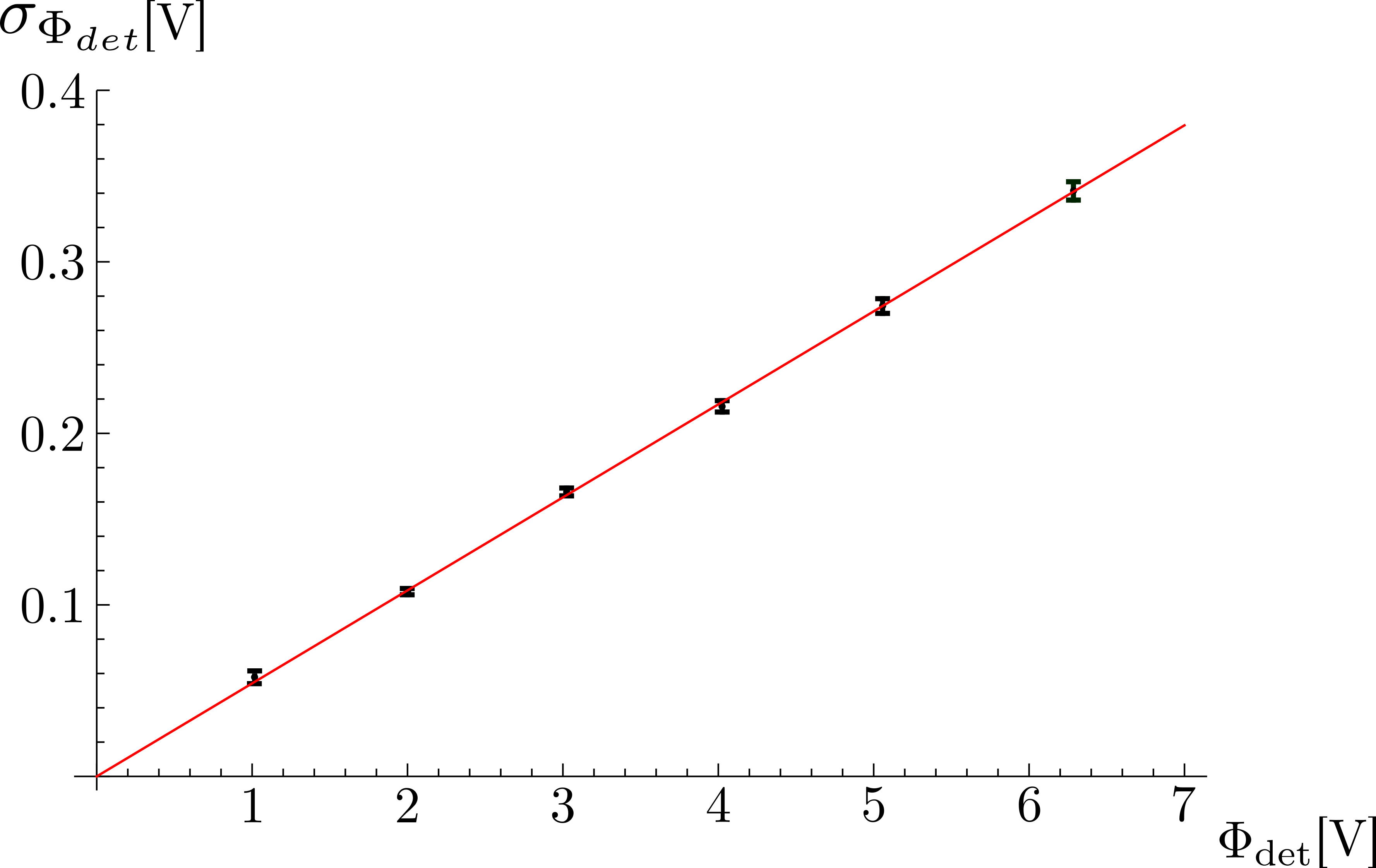}
\caption{\textbf{Power dependence of detector sensitivity.} $\sigma_{\phi_{det}}$ as a function of the power in the detector show a linear response. The optical power is attenuated right in front of the detector, for a fixed trap frequency of $\omega_0$}
\label{fig:calibration1}
\end{figure}

The particle position is measured using a standard split-mirror detection scheme: The forward propagating \tzz light mode is separated from the \tzo mode, attenuated and split along the vertical axis by means of a sharp-edged D-shaped mirror. Light is attenuated and sent to a pair of balanced InGaAs detectors (PDB425C, Thorlabs Inc.) where the difference signal $\phi_{det}$ is measured. The displacement of the particle is proportional to that signal when the particle is in trapped close to the center of the potential ($\Delta x= c_{center}\phi_{det}$). 
In this condition, the calibration factor $c_{center}$ can be calculated for each trapped particle using an independent measurement, where the power of the \tzo mode is set to zero. By assuming harmonic motion in the \tzz, we use the equipartition theorem: $m\omega_0^2 c_{center} \sigma_{\phi_{det}}(\omega_0)^2=kT$. Here $m=(2.831\pm0.001) \cdot 10^{-18} \text{kg}$ is the mass of the silica nanosphere as provided by the manufacturer specifications, $k$ is the Boltzmann constant, $T=300$K is the room temperature and $\omega_0=2\pi(210.1\pm 0.8)\text{kHz}$ is the particle frequency, for a power in the optical trap of $P_0=(100\pm1)\text{mW}$, where $\sigma_{\phi_{det}}(\omega_0 ) =(0.042\pm0.001)\text{V}$ is the standard deviation of the detector signal. In Fig. \ref{fig:calibration1} we show a typical dependence of $\sigma_{\phi_{det}}(\omega_0)$ on the power at the detector, where the particle trapped at a pressure 0.1 \text{mbar} and only in the \tzz mode at a constant power, ($\alpha=1,\beta=0,F=0$). For this configuration, we obtain:
$$
c_{center}=\sqrt{\frac{kT}{m}}  \frac{1}{\sigma_{\phi_{det}} \omega_0}=(6.95\pm 0.06)\cdot 10^{-7} m/V.
$$

However, when the particle is significantly off-center (for example when the particle is in either of the two wells of the \tzo mode) the sensitivity (i.e. the calibration factor) of the detection scheme is reduced, because less light from the \tzz mode is scattered by the particle. In fact, if we repeat the above calibration for when the particle is on the left and right well, we get:
$$
\begin{array}{cc}
c_{left}&=(2.95\pm 0.04)\cdot 10^{-6} m/V. \\
c_{right}&=(2.62\pm 0.03)\cdot 10^{-6} m/V. 
\end{array}
$$
Thus, we investigate the dependence of the calibration factor on the particle position $c(x)$ for the relevant spatial interval. This is possible by an analysis of the particle dynamics in an equilibrium scenario, as will be discussed in the following.
The data used to calibrate the effect stems from the re-initialisation part of the protocols, i.e. is taken independently of the erasure protocol and with precisely the same experimental configuration and particle. During the re-initialisation phase we obtain trajectories with the particle equilibrated in each of both wells, respectively. We obtain this data for two different powers, with the potential set to "high" and "low". We label $s(t)$ the uncalibrated particle trajectory registered by the detector. The goal is to reconstruct calibrated particle trajectory in real space $x(t)$. We define $\eta(s)=dx/ds$ as the differential calibration factor, which we expect to correspond to the calibration factor $c_i$ as discussed above (i= left, right, center). Accordingly, $\partial V/\partial s = \partial V/\partial x \cdot \partial x/\partial s = F(x) \eta(s)$. 
 Let us first consider an isolated particle (no friction, no force noise) moving in a potential landscape and detected in a scheme with a position dependent calibration factor:
\begin{eqnarray}
F(x) &=& \frac{1}{\eta(s)} \cdot \frac{\partial V}{\partial s} \\
F(s(x))/m &=& \frac{\partial}{\partial t}(\dot{x}) =
\frac{\partial}{\partial t}(\eta(s)\dot{s})=
\frac{\partial \eta}{\partial s} \dot{s}^2 + \eta(s) \ddot{s}\\
 &=& \frac{1}{\eta(s)}\left(\frac{\partial \eta^2}{\partial s}\cdot \frac{\dot{s}^2}{2} +\eta^2 \ddot{s} \right) 
\end{eqnarray}
From this we can deduce how Newtons laws are transformed for the uncalibrated trajectories in dependence of the differential calibration factor. 
\begin{equation}
    \frac{\partial V}{\partial s} = \frac{\partial \eta(s)^2}{\partial s}\cdot \frac{\dot{s}^2}{2}+m\eta^2(s)\ddot{s}
\end{equation}
Clearly, the detected uncalibrated trajectory allows to determine the calibration factor using the kinetic quantities in this equation. Consider a particle that is weakly coupled to and in equilibrium with an environment. For a sufficiently large ensemble of data where a particle crosses a specific position, the above equations hold on average, as random force noise as well as friction cancel on average.
Accordingly:
\begin{equation}
    \langle \frac{\partial V}{\partial s} \rangle = \frac{\partial \eta(s)^2} {\partial s} \cdot \langle \frac{\dot{s}^2}{2} \rangle +m\eta^2(s) \langle \ddot{s} \rangle
\end{equation}
 The calibration factor is now expressed in terms of a first order differential equation with uncalibrated quantities, which are known: the reconstructed potential derivative, the average kinetic energy and the average acceleration when the particle is in the \tzz mode and in the wells of the \tzo high and low power mode. We numerically solve the differential equation to reconstruct $\sqrt{\eta^2(s)}$.  
 
\begin{figure*}[h!tb]
\centering
\includegraphics[width=0.90\textwidth]{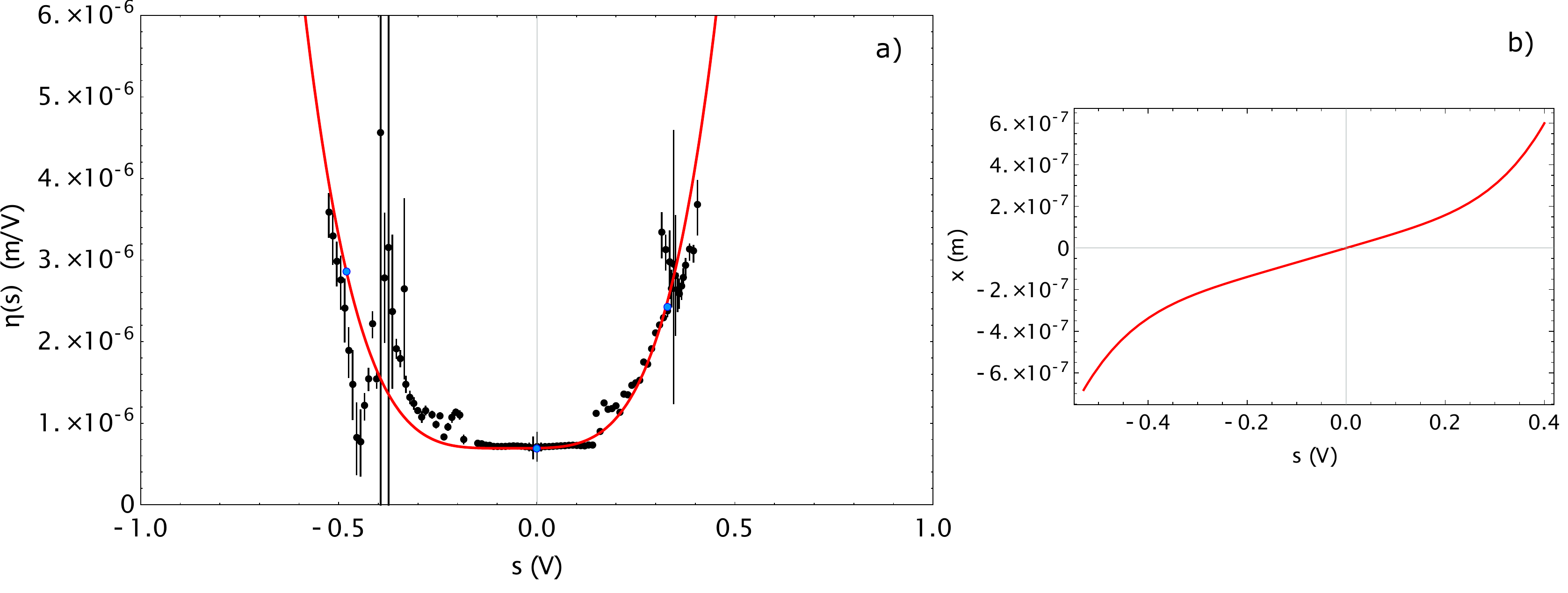}
\caption{\textbf{Detector Sensitivity.} a) Differential calibration factor as a function of the detector response. Black points: experimental values. Error bars are 1-sigma, evaluated with a Montecarlo simulation. Blue points: calibration factor evaluated using the equipartation theorem in approximately harmonic traps with error bars are smaller than point size. Red: fit of all data points with a quartic function. b) Correspondingly calibrated position as a function of the detector response (uncalibrated position).}
\label{fig:calibration_2021}
\end{figure*}

In Fig. \ref{fig:calibration_2021}a we show the differential calibration factor as a function of the voltage response of the detector. The data is obtained by stitching the numerical results for the "high" and "low" as well as the \tzz potential. While this allows to cover the full relevant range of data, it results in two apparent discontinuities and corresponding large error bars due to numerical mismatch in the different reconstructions. In fact for each component we manually adjust the position offset $\langle V(s+\epsilon=\epsilon_0) \rangle$ to minimize discontinuities. We evaluate error bars on the points by performing a Montecarlo simulation where we randomly generate $\epsilon$ in an around of $\epsilon_0$. In order to compare this dynamics-based full-range calibration method with the standard measurement for harmonic potentials described before, Fig.\ref{fig:calibration_2021} also features the values  $c_{left}$ and $c_{right}$ and $c_{center}$ as blue dots showing a good agreement between the two calibration methods.

%It is created by joining the numerical results for the different parts of the potential. The error bars, and the apparent discontinuities are given by offset between the reconstructed potential and the kinetic terms in the differential equation. Indeed "zeroes" in these functions have to match to avoid numerical discontinuities in the solution. For each, sweep, thus, we manually shift $\langle U(s+\epsilon=\epsilon_0) \rangle$ to avoid discontinuities. We also evaluate error bars on the points by performing a Montecarlo simulation where we randomly generate $\epsilon$ in an around of $\epsilon_0$. In Fig.\ref{fig:calibration_2021} we also plot the values of $c_{left}$ and $c_{right}$ and $c_{center}$ as blue dots showing a good match between the two calibration methods. 

We interpolate the data points with the function $\alpha_4 (s-s_0)^4+\alpha_0$, which preserves the expected flatness of the calibration function around zero. In Fig. \ref{fig:calibration_2021}b we show the integrated calibration factor $\eta_0(x)=\int_{s_{min}}^x \eta(s')ds'$, which allows us to determine the real space trajectory for any of our measurements in  post processing.

%Additionally, in our position measurement, the signal amplitude depends on the incident power on the detector. We verify this dependence by changing the attenuation of the light on the detector, without changing the power in the trap. In Fig. \ref{fig:calibration1} we prove that the relation is linear, and from a fit we extrapolate the generalized calibration factor, now independent from the power at the detector. %We obtain $c_{det}=0.054\pm0.001$. 
Every measured timetrace is rescaled to account for power fluctuations, which are monitored during the whole measurement, and their linear effect on the calibration in Fig. \ref{fig:calibration1}. Specifically, the data is normalised to a power $P_0$ corresponding to a measured voltage of $\phi_{det_{max}}=(5.51\pm0.02)V$. 

%Note that this correction does not only affect the measured amplitude of motion, but also the signal offset.  

%The measurements of the particle position performed during a protocol are thus normalized to a calibrated value. This is realized by applying a correction function on the detector signal which is dependent on the inverse of the power $P_0$, independently measured during each protocol, and normalized to $\phi_{det_{max}}=(5.51\pm0.02)V$. Note that this correction does not only affect the measured amplitude of motion, but also the signal offset.

\subsection{Electrodes Calibration}
\begin{figure}[h!tb]
\centering
\includegraphics[width=0.90\columnwidth]{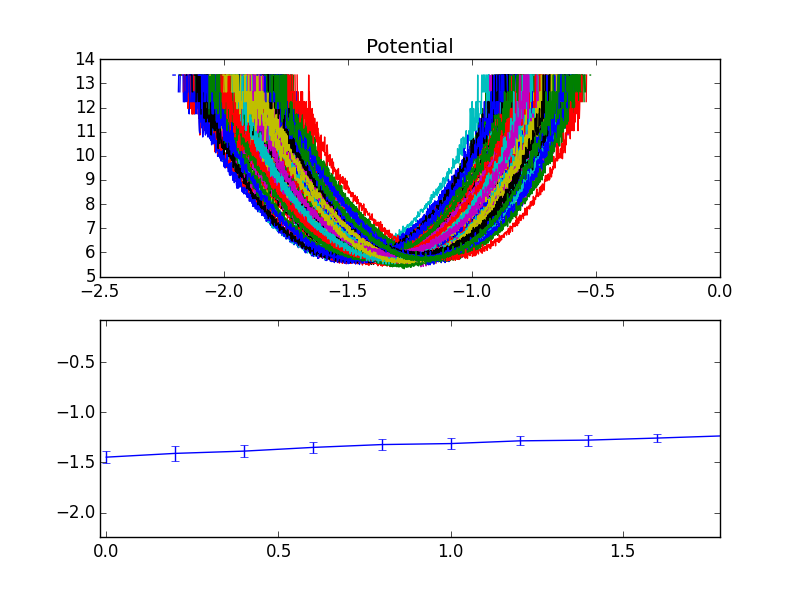}
\caption{\textbf{Calibration of the electrodes.} Top) Potential reconstruction of a particle in a \tzz mode for various values of the electrodes voltage. Bottom) Average position of particle as a function of electrodes voltage.}
\label{fig:calibration2}
\end{figure}

We calibrate the displacement force $F=c_{f}\phi_f$ due to the voltage $\phi_f$ applied to the electrodes by measuring the displacement of the particle in the \tzz mode as a function of the voltage applied to the electrodes. The electrodes voltage is increased from $\phi_f=-7.5$ V to $\phi_f=7.5$ V in multiple steps using a voltage amplifier. At every step, we measure the mean position of the particle in the tweezer. A repetition of at least 10 times decreases the statistical error and compensates for position drifts. Afterwards, the measurement is fitted with a linear function (see Fig. \ref{fig:calibration2}) from which we extract the parameter $c_{f}$. We get:
$$
 \frac{\langle \phi_{det} \rangle}{\phi_f}= \frac{\langle x \rangle /c_{det}}{F/c_f}=\frac{1}{m\omega_0^2}\cdot \frac{c_{det}}{c_f}
$$
Note: every particle has a different number of charges, and thus needs a different calibration. All protocols presented here are taken for the same particle. The force calibration factor determined for this particle is $c_{f}=(5.1\pm 0.3) 10^{-14} N/V$. The maximal tilting force applied in the protocol is $F_{max}=4.4\cdot 10^{-14} \text{N}$ and corresponds to  $\phi_{f_{max}}= F_{max}/c_{f}= 0.86 V$ .

\section{Potential Model}
\begin{figure*}[h!tb]
\centering
\includegraphics[width=0.90\textwidth]{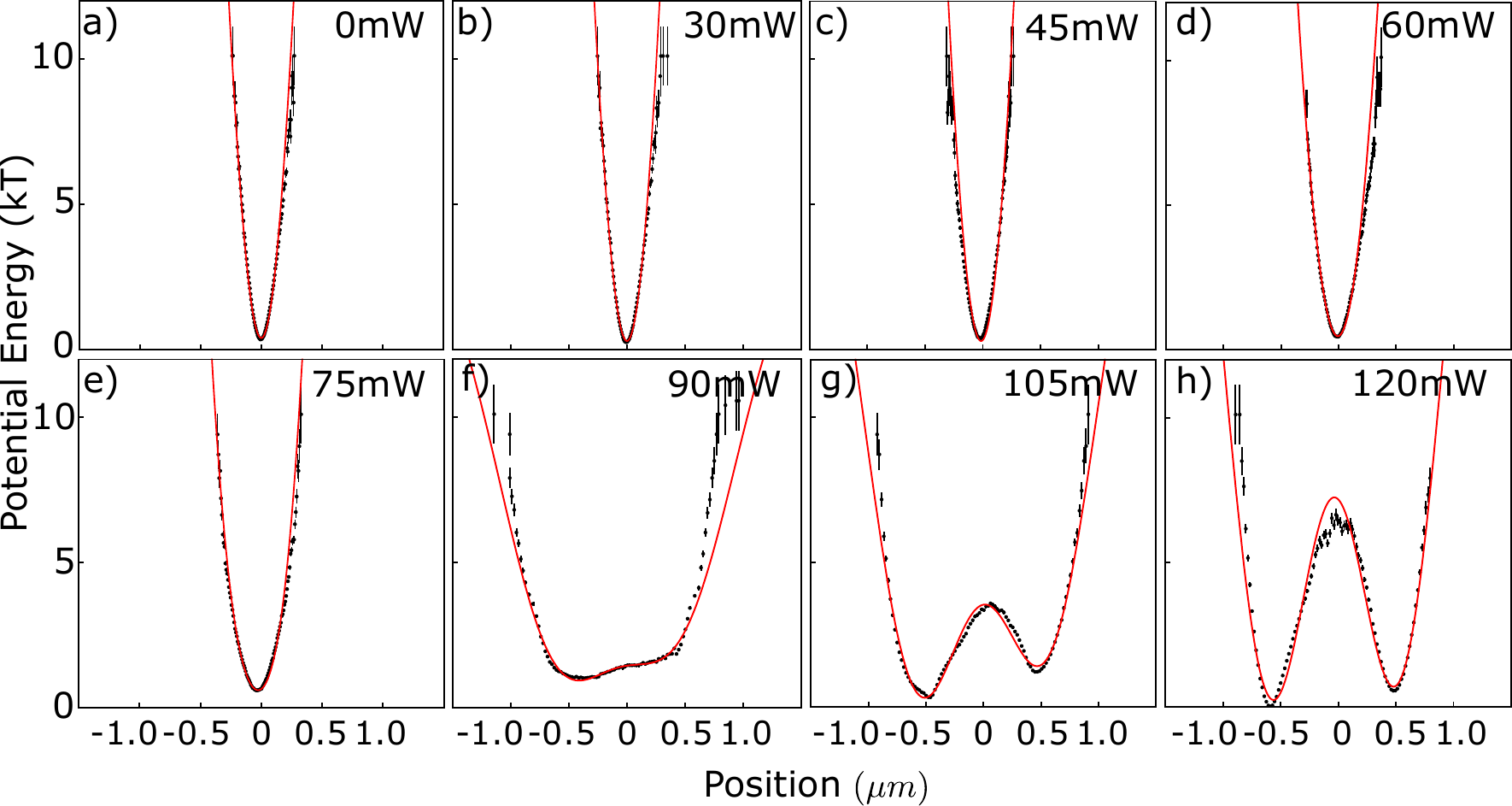}
\caption{\textbf{Potential reconstruction.} Reconstruction of U(x), fixing the power of the \tzz after the vacuum chamber to $\alpha_0$=50mW, and varying the power of \tzo from $\beta=120mW$ (h) to $\beta=0mW$ (a).}
\label{fig:potentialtime}
\end{figure*}

The calculation of work applied during a protocol is based on the knowledge of the optical potential at every time. We model the full time-dependence based on our time-dependent parameter-settings and the reconstuction of the potential for relevant, specific parameter settings, corresponding to snapshots of the protocol. The potential reconstruction for these snapshots is obtained from a series of independent measurements: we first let the particle equilibrate to the respective potential landscape. Then we measure the particle’s position probability distribution $p(x)$ taking a timetrace which is sufficiently long ($100$~ms) to safely use the ergodic assumption. The potential is then reconstructed by using: $V(x)=-kT\ln(p(x))$ with Boltzmann constant $k$, temperature T. 
We measure the potential for various powers in \tzo and fixed \tzz power. Measurements are then sensitivity corrected, calibrated (see III,A), and fitted to find a model for the potential depending on the power of \tzz and \tzo (see some examples of measured potentials in Fig. \ref{fig:potentialtime}). Every potential is fitted with a function which approximates the optical potential as a combination of \tzz and \tzo modes:

$f(x)=c-\frac{8 \alpha  (x-\text{x0})^2 e^{-\frac{2 (x-\text{x0})^2}{\sigma ^2}}}{\pi  \sigma ^6}-\frac{2 \beta  e^{-\frac{2 (x-\text{x1})^2}{\sigma ^2}}}{\pi  \sigma ^4}$

We first identify parameter $\beta$ by fitting our potential when \tzo is off (Fig. \ref{fig:potentialtime}a). We use a “jumping” potential (i.e., when the particle jumps between the wells, Fig. \ref{fig:potentialtime}h) to infer $|x_0-x_1|$, which identifies the asymmetry in our model, which we assume constant for the whole experiment. We then fit the remaining potentials to find and interpolate parameters $\alpha$, $x_0$, $\sigma$ and $c$ as a function power in the mode \tzo. This process gives us the full ability to program $V(x,t)$ for our protocol via the input parameters.

\section{Protocols and equilibrium phase space}

\begin{figure}[h!tb]
\centering
\includegraphics[width=0.90\columnwidth]{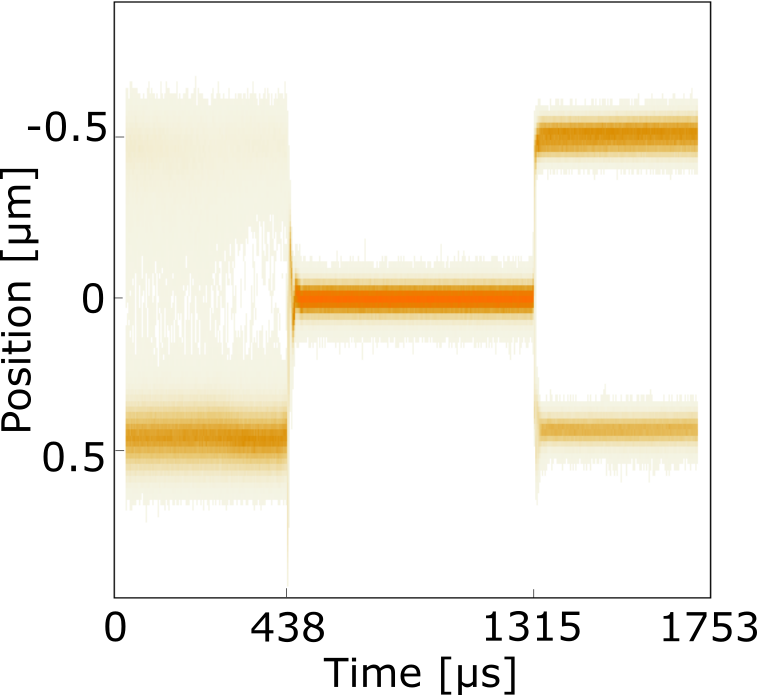}
\caption{\textbf{Probability Density of the reinitialisation part of the protocol.} Average position over $\approx 10000$ protocol repetitions. In the first part the particle is in the "low" potential, after the reset protocol. At $t=438\mu s$ the power of \tzz is maximised. After $t=1315 \mu s$  the \tzo power is also maximised. At the end of the reinitialisation, the particle is reequilibrated to a "high" power potential.}
\label{fig:reset01}
\end{figure}

\begin{figure}[h!tb]
\centering
\includegraphics[width=0.90\columnwidth]{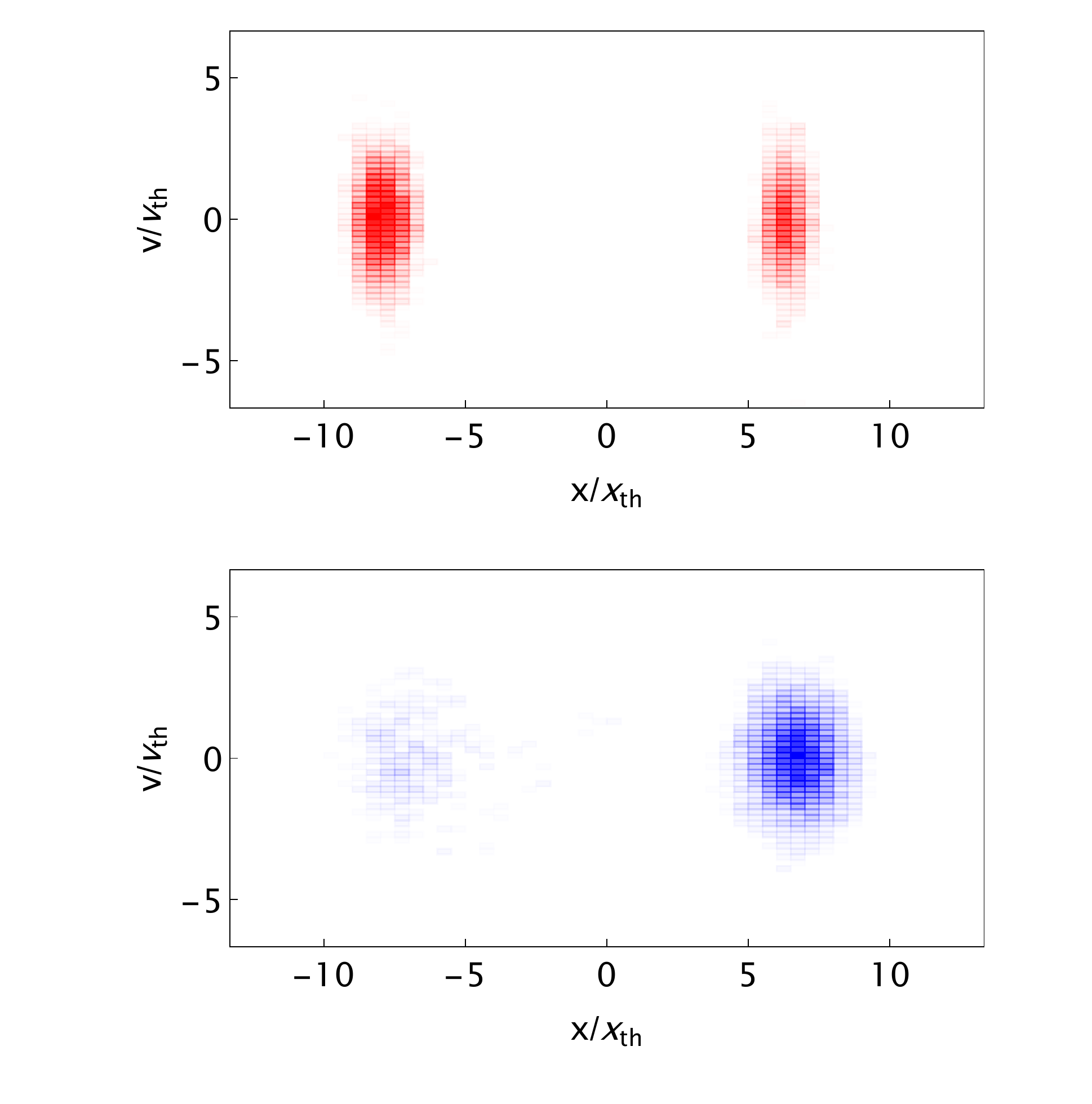}
\caption{\textbf{x-v phase-space of the erasure protocol.} Top) Initial state (out of equilibrium) state, Bottom) Final state after reset. Position and velocity are normalised to $x_{th}=\sqrt{k_B T/(m(2\pi\nu_0)^2}$ and $v_{th}=\sqrt{k_B T/m}$, where $\nu=80kHz$ and $m=2.83fg$. }
\label{fig:phasespace}
\end{figure}

Every protocol timetrace is measured using a split detection scheme, and sampled at $6.25 MS/s$.  A calibrated timetrace $x^l (t_i)$ (i is the i-th time step, l is the l-th repetition of the protocol). A full protocol consists of three parts. In the first part we perform a fast ($t_{up}=3\mu s$) increase in the power of the trap, followed by a slower ($t_{neq}=358\mu s$) decrease back to the memory potential; % The particle starts in equilibrium with the steep potential, $( x(t_0 )\in \rho^{neq}(x) )$, with probability $p_0^{eq}$ of being in the left well. At first, the potential is rapidly changed by $\alpha = 1 \rightarrow \epsilon+1$ brought from shallow to steep ($t_{up}=3\mu s$), then slowly brought down in a linear fashion to the shallow one $t_{neq}=358\mu s$. 
In the second part is the equilibrium erasure, where \tzo is quasi-adiabatically modified to first decrease and then increase again the potential barrier ($t_{eq}=1358\mu s$). At $t_1=728\mu s$ we introduce a linear increase of the electrostatic force that reaches its maximum $F_{max}$ at $t_2=1258\mu s$, decreasing again linearly till the end of the protocol.  The third part is a reset protocol that restarts the particle in its original (non-equilibrium) distribution $\rho_{neq}$. This part is not evaluated for the work and heat consumption. First the particle is let to equilibrate to the final position, then \tzz is rapidly increased to its maximum value. Finally, \tzo is also increased rapidly. This sets the particle to either of the two wells with a probability that depends on the symmetry of the potential. Finally, the particle is let to equilibrate to this potential. We show in Fig. \ref{fig:reset01} the probability density during the reinitialisation part of the protocol for the $\epsilon=4$ data-set. Overall, the entire data-set lasts $t_{tot}=3116\mu s$. The double-well potential is consistently set to the same power ratio between the two modes (\tzo is 2.7 times the one of the \tzz measured right after the vacuum chamber). Parameter $\epsilon$ is set such as the trap power is $(\epsilon+1)$ times stronger than the power for the equilibrium potential. In Fig. \ref{fig:phasespace} we show the phase-space distribution of the particle state at the beginning (top) and at the end of the erasure protocol (bottom).

\section{Evaluation of Landauer's bound}
Given a protocol timetrace  $x^l (t_i)$ and the potential model $V(x,t)$, we define the average dissipated work as:

\begin{align*}
\langle W \rangle_\epsilon & = \langle \int_0^{t_{eq}} \frac{\partial{V(x,t)}}{\partial t} dt\rangle_\epsilon \rightarrow \\
&\rightarrow \langle \sum_{i=0}^{t_{fin}} \left[ V(x_i,t_{i+1} )-V(x_i,t_i)\right] \rangle_\epsilon
\end{align*}
The average dissipated heat during the protocol, following the First Law is:

\begin{align*}
\langle Q\rangle_\epsilon & = \langle W \rangle - \langle \Delta E \rangle
\end{align*}

Here  $\langle \rangle$ is the average over $M\approx 10000$ protocols, and $\epsilon$ is the non-equilibrium parameter. In addition, we define $\Delta E_i = E_i (t_{fin})-E_i (t_0) = \frac{1}{2}m((v_{fin}^i)^2-(v_{0}^i)^2)+[V(x_{fin}^i,t_{protocol})-V(x_{init}^i,t_{protocol})]$ as the total energy variation between the beginning (out-of-equilibrium) and ending (equilibrium) of the protocol. In Table \ref{tab:data} we report experimental values for average work, heat, and energy difference as a function of $\epsilon$. Uncertainty is evaluated through the standard deviation of the resulting distribution, and propagated for the averaging over all the repetitions of the protocol. %In Fig. \ref{} we plot the work distributions for different values of $\epsilon$, showing a decrease in their average for growing $\epsilon$. 

\begin{widetext}
\begin{center}
\begin{table}
\begin{tabular}{|c|c|c|c|c|c|c| } 
 \hline
 $\epsilon$ & $p_0^{eq}$ & $p_0^{f}$ & $\langle W \rangle$ & $\langle \Delta E \rangle$ & $\langle Q \rangle$ & $\Delta W_{asym}$ \\ 
 0 & $0.634\pm0.006$ & $0.040\pm0.001$ & $0.583\pm0.045$ & 0 & $0.62\pm0.064$ & $0.121\pm 0.047$ \\ 
 1 & $0.415\pm0.006$ & $0.004\pm0.001$ & $0.487\pm0.048$ & $0.185\pm0.031$ & $0.295\pm0.06$ & $0.060\pm 0.016$ \\
 2 & $0.423\pm0.007$ & $0.0046\pm0.0014$ & $0.351\pm0.046$ & $0.283\pm0.070$ & $0.024\pm 0.086$ & $0.081\pm0.019$ \\
 3 & $0.637\pm0.021$ & $0.033\pm 0.010$ & $0.196\pm0.042$ & $0.447\pm0.042$ & $-0.215\pm0.06$ & $0.156\pm0.070$ \\
 4 & $0.566\pm0.015$ & $0.0210\pm0.0035$ & $0.054\pm0.023$ & $0.446\pm0.023$ & $-0.328\pm0.032$ & $0.222\pm0.028$ \\
 \hline
 \end{tabular}
\caption{\label{tab:data}\textbf{Experimental parameters.} $\epsilon$ is the non-equilibrium parameter. $p_0^{eq}$ is the probability of the particle starting in well 0. $p_0^{f}$ is the probability of the particle being in well 0 after the protocol. $\langle W\rangle$ is the average extracted work during all the repetitions of the protocol. $\langle \Delta E\rangle$ is the average energy difference between start and end of the protocols. $\langle Q \rangle $ is the average dissipated heat during the protocols (a negative value indicates cooling of the environment).  $\langle \Delta W_{asym} \rangle $ is the correction to Landauer's bound given by the asymmetry of the potential and non-perfect erasure.}
\end{table}
\end{center}
\end{widetext}